\documentclass[aps,twocolumn,superscriptaddress,prl]{revtex4-1}
\usepackage{graphicx}
\usepackage{amsmath}
\usepackage{nccmath}
\usepackage{txfonts}
\usepackage{bm} 
\usepackage[hypertex,colorlinks=true,linkcolor=black,citecolor=blue,filecolor=blue,urlcolor=blue,setpagesize=false,nesting=true]{hyperref}

\newcommand{\msr}{$\mu$SR}
\newcommand{\lvo}{LiV$_2$O$_4$}
\newcommand{\lzvo}{Li$_{1-x}$Zn$_x$V$_2$O$_4$}
\newcommand{\ymn}{Y(Sc)Mn$_2$}

\begin{document}

\preprint{arXiv}

\title{Metallic Spin Liquid-like Behavior of \lvo}

\author{H. Okabe}
\affiliation{Muon Science Laboratory, Institute of Materials Structure Science, High Energy Accelerator Research Organization (KEK), Tsukuba, Ibaraki 305-0801, Japan}
\affiliation{Condensed Matter Research Center, Institute of Materials Structure Science, High Energy Accelerator Research Organization (KEK), Tsukuba, Ibaraki 305-0801, Japan}
\affiliation{The Graduate University for Advanced Studies (Sokendai), Tsukuba, Ibaraki 305-0801, Japan}
\author{M. Hiraishi}
\affiliation{Muon Science Laboratory, Institute of Materials Structure Science, High Energy Accelerator Research Organization (KEK), Tsukuba, Ibaraki 305-0801, Japan}
\affiliation{Condensed Matter Research Center, Institute of Materials Structure Science, High Energy Accelerator Research Organization (KEK), Tsukuba, Ibaraki 305-0801, Japan}
\author{A. Koda}
\affiliation{Muon Science Laboratory, Institute of Materials Structure Science, High Energy Accelerator Research Organization (KEK), Tsukuba, Ibaraki 305-0801, Japan}
\affiliation{Condensed Matter Research Center, Institute of Materials Structure Science, High Energy Accelerator Research Organization (KEK), Tsukuba, Ibaraki 305-0801, Japan}
\affiliation{The Graduate University for Advanced Studies (Sokendai), Tsukuba, Ibaraki 305-0801, Japan}
\author{K. M. Kojima}
\affiliation{Muon Science Laboratory, Institute of Materials Structure Science, High Energy Accelerator Research Organization (KEK), Tsukuba, Ibaraki 305-0801, Japan}
\affiliation{Condensed Matter Research Center, Institute of Materials Structure Science, High Energy Accelerator Research Organization (KEK), Tsukuba, Ibaraki 305-0801, Japan}
\affiliation{The Graduate University for Advanced Studies (Sokendai), Tsukuba, Ibaraki 305-0801, Japan}
\author{S. Takeshita}
\affiliation{Muon Science Laboratory, Institute of Materials Structure Science, High Energy Accelerator Research Organization (KEK), Tsukuba, Ibaraki 305-0801, Japan}
\affiliation{Condensed Matter Research Center, Institute of Materials Structure Science, High Energy Accelerator Research Organization (KEK), Tsukuba, Ibaraki 305-0801, Japan}
\author{I. Yamauchi}
\affiliation{Department of Physics, Graduate School of Science and Engineering, Saga University, Saga 840-8502, Japan}
\author{Y. Matsushita}
\affiliation{National Institute for Materials Science (NIMS), Tsukuba, Ibaraki 305-0044, Japan}
\author{Y. Kuramoto}
\affiliation{Condensed Matter Research Center, Institute of Materials Structure Science, High Energy Accelerator Research Organization (KEK), Tsukuba, Ibaraki 305-0801, Japan}
\affiliation{Department of Physics, Kobe University, Kobe 657-8501, Japan}
\author{R. Kadono}\email{Corresponding author: ryouke.kadono@kek.jp}
\affiliation{Muon Science Laboratory, Institute of Materials Structure Science, High Energy Accelerator Research Organization (KEK), Tsukuba, Ibaraki 305-0801, Japan}
\affiliation{Condensed Matter Research Center, Institute of Materials Structure Science, High Energy Accelerator Research Organization (KEK), Tsukuba, Ibaraki 305-0801, Japan}
\affiliation{The Graduate University for Advanced Studies (Sokendai), Tsukuba, Ibaraki 305-0801, Japan}

\begin{abstract}
\lvo\ spinel is known to exhibit heavy fermion-like behavior below a characteristic temperature $T_K\simeq 20$ K, while it preserves a paramagnetic state down to $T\sim10^{-2}$ K due to geometrical frustration. Here, it is shown that the dynamical spin susceptibility $\chi({\bf q},\omega)$ in \lvo\ exhibits anomalous duality which is modeled as a sum of itinerant ($\chi_{\rm F}$) and local ($\chi_{\rm L}$) components,and that the local spin dynamics inferred from $\chi_{\rm L}({\bf q},\omega)$ is qualitatively different from that expected from time-averaged bulk properties.
The anomaly coexists with the marginal Fermi liquid behavior inferred from the $-\ln T$ dependence of the electronic specific heat over a wide temperature range below $T_K$. We argue that such unusual properties of \lvo\ can be attributed to the putative {\it metallic spin liquid} state emerging near the quantum critical point between spin glass and Fermi liquid states.
\end{abstract}


\maketitle

Quantum phase transition and associated critical behavior of electronic states has been a central focus of condensed matter physics in the past decades \cite{Gegenwart:08}.  The quantum fluctuation induces various anomalies to the electronic properties at finite temperatures, serving as a promising ground in hunting for novel states of matter. In particular, the ``metallic spin liquid" state (or spin liquid metal) is attracting much interest as a novel non-Fermi liquid state in the field of $f$-electron systems, where it is predicted to emerge near the quantum critical point (QCP) next to the metallic spin glass state \cite{Sachdev:93,Sachdev:96,Burdin:02,Custers:10}.  

The metallic spin liquid comprises a counterpart of the ``spin liquid" in insulators. While the spin liquid is characterized by disappearance of paramagnetism (as local electron spins fall into a collective singlet ground state), the metallic spin liquid exhibits paramagnetism linked to spin glass. The ultimate conflict between the strong electronic correlation (preferring magnetic order) and the Kondo effect (driving to the Fermi liquid) may lead to quantum criticality and associated novel metallic state, where the coupling between spin and charge fluctuation is largely different from the insulating spin liquid \cite{Sachdev:96,Burdin:02}.  It is also noticeable that the metallic spin glass/liquid is intensively discussed as a stage of applying the so-called anti-de-Sitter/conformal field theory (AdS/CFT) correspondence developed in the quantum field theory of gravity and entropy of black holes \cite{Sachdev:10,Maldacena:16}.

It has been recently demonstrated in some of the $f$-electron antiferromagnetic (AF) metals with pyrochlore \cite{Nakatsuji:06} or Kagome lattice structure \cite{Fritsch:14} that the suppression of magnetic order by geometrical flrustration leads to non-Fermi liquid behavior emerging in their bulk electronic properties such as resistivity ($\rho$), magnetic susceptibility ($\chi_{\rm bulk}$), and electronic specific heat coefficient ($\gamma$). In this regard, it is noteworthy that the heavy fermion-like $d$-electron compounds known to date, i.e., \ymn\ \cite{Wada:89,Fisher:93} and \lvo\ \cite{Kondo:97,Urano:00,Matsushita:05} have a common feature that the transition metal ions comprise the pyrochlore lattice and therefore subject to the geometrical frustration, as inferred from the emergence of metallic spin glass upon chemical pressure \cite{Mekata:97,Ueda:97,Trinkl:00,Koda:04}.  In fact, alienation from the Fermi liquid was indeed suggested by ``$-\ln T$" dependence of $\gamma$  (i.e., the {\sl marginal} Fermi liquid behavior regarding the conduction electrons) and anomalous magnetic field dependence of $^7$Li-NMR for powder samples of \lvo\ \cite{Kaps:01,Buttgen:10}.

This motivates us to reexamine the currently prevailing consensus of \lvo\ as a typical heavy fermion (HF) metal from the viewpoint of metallic spin glass/liquid.  While certain bulk properties ($\rho$, $\chi_{\rm bulk}$) and  the Korringa law indicated by $^7$Li-NMR suggest Fermi liquid state \cite{Kondo:97,Urano:00,Matsushita:05,Mahajan:98,Johnston:05,Shimizu:12},  muon spin rotation (\msr) \cite{Koda:04,Koda:05},  $^{51}$V-NMR \cite{Shimizu:12,Takeda:15}, and inelastic neutron scattering (INS) \cite{Lee:01,Murani:04, Tomiyasu:14}  coherently imply presence of localized $d$ electrons even below a characteristic temperature $T_K\simeq20$ K where the $-\ln T$ behavior for $\gamma$ develops.  New theoretical framework beyond the canonical Kondo lattice model is called for, since inter-site interactions including the Coulomb and AF correlations clearly pertains to the anomaly \cite{Burdin:02,Laad:03}.  

We report \msr\ study on high-quality \lvo\ samples that provides crucial information on the dynamical spin susceptibility [$\chi({\bf q},\omega)=\chi'({\bf q},\omega)+i\chi''({\bf q},\omega)$]. The muon Knight shift ($K_{\mu}$) and longitudinal depolarization rate ($1/T^{\mu}_1$) were measured simultaneously on the same sample to entirely eliminate the sample dependence and aging problem. Appropriate choice of external magnetic field ($B_0\le0.5$ T) allowed determination of $K_{\mu}$ with improved precision by controlling the transverse linewidth  ($1/T^{\mu}_2$). 
We show that both $K_{\mu}$ and $1/T^{\mu}_1$ are dominated by paramagnetism, and that $1/T^{\mu}_2$ is anomalously enhanced by $B_0$. These features are commonly observed in the paramagnetic state of canonical dilute spin glass systems like {\it Ag}Mn \cite{Heffner:82}, which is in marked contrast with the HF-like properties. We further demonstrate that such dichotomy is understood by a phenomenological model in which $\chi({\bf q},\omega)$ is described by a sum of itinerant ($\chi_{\rm F}$) and local ($\chi_{\rm L}$) spin components, to which the relevant probes exhibit complementary sensitivity. The behavior of $\chi_{\rm L}$ inferred from $K_{\mu}$ and $1/T^{\mu}_1$ is qualitatively in line with that of the local susceptibility predicted for the metallic spin liquid \cite{Burdin:02}.

\begin{figure}[t]
\begin{center}
\includegraphics[width=0.95\linewidth]{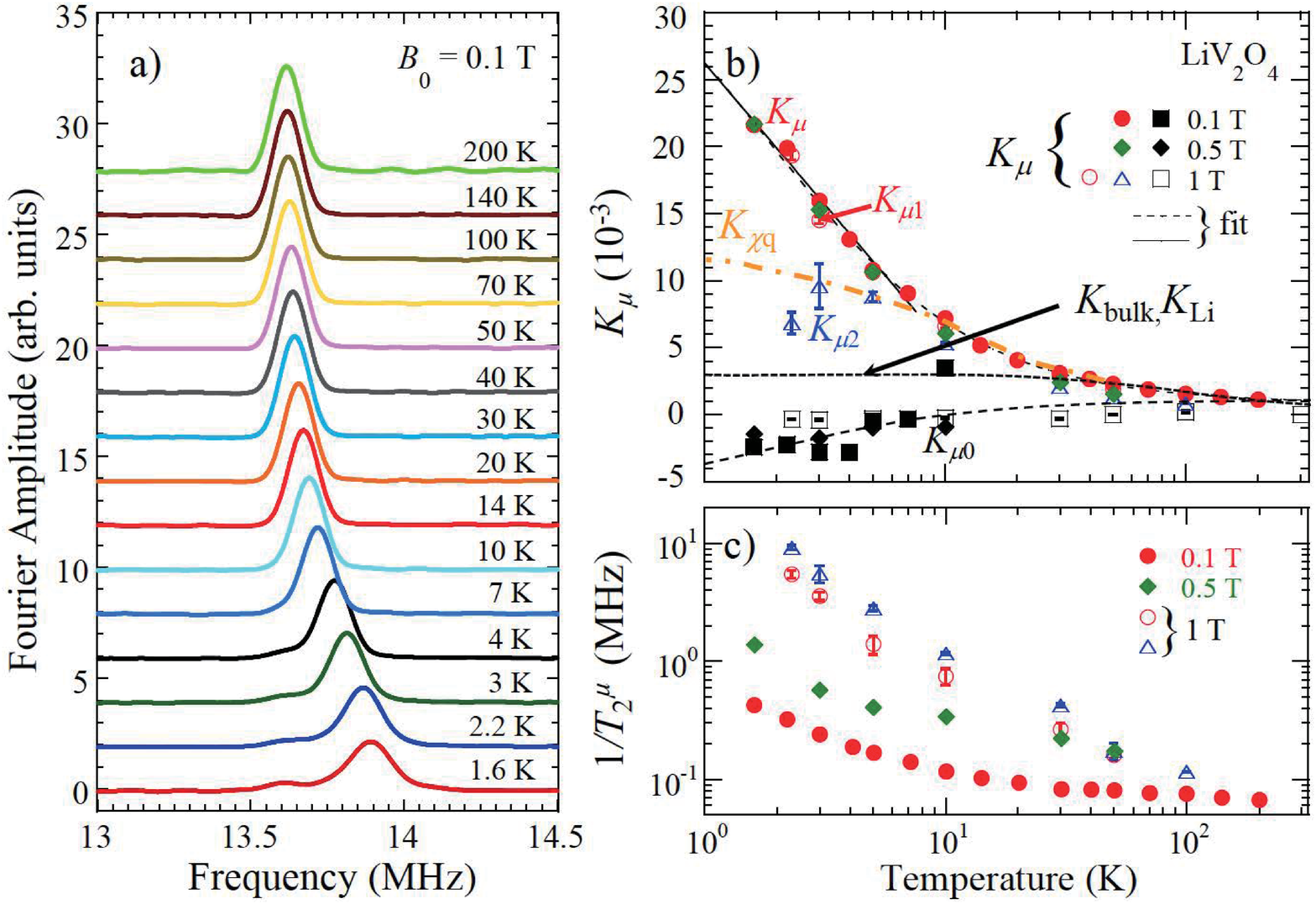}
\caption{(Color online)
(a) Fourier transform of the \msr\ time spectra under a transverse field of 0.1 T (real amplitudes) at various temperatures in \lvo. (b) Muon Knight shift versus temperature at 0.1 and 0.5 T (solid symbols), where the shift splits into the primary ($K_{\mu}$) and satellite ($K_{\mu0}$) components below $\sim$10$^1$ K. (c) Transverse linewidth ($1/T_2^\mu$) for the $K_{\mu}$ component in (b). Open symbols in (b) and (c) show previous result at 1 T described by two components ($K_{\mu1}$ and $K_{\mu2}$) \cite{Koda:05}. The $^7$Li-NMR data \cite{Shimizu:12} and those evaluated respectively from bulk susceptibility \cite{Matsushita:05}  ($K_{\rm Li}$ and $K_{\rm bulk}$, dashed curve), and static susceptibility deduced from INS experiment \cite{Lee:01} ($K_{\chi {\rm q}}$, dot-dashed curve) are also shown in (b) for comparison. For the fit curves in (b), see text. }
\label{KL}
\end{center}
\end{figure}

\begin{figure}[t]
\begin{center}
\includegraphics[width=1.0\linewidth]{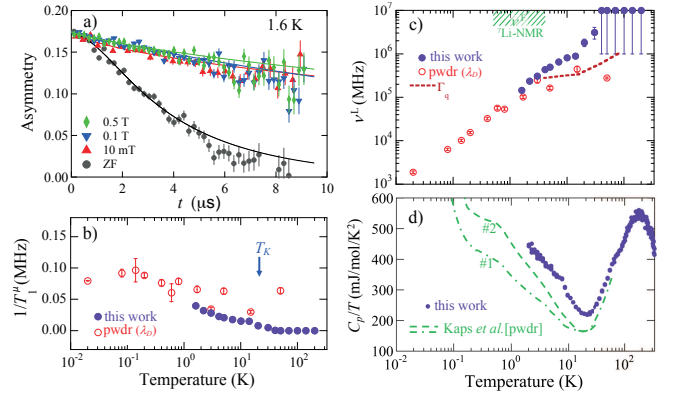}
\caption{(Color online)
(a) \msr\ time spectra under a longitudinal field (LF) of 0, 0.01, 0.1 and 0.5 T observed at 1.6 K. Temperature ($T$) dependence of (b) longitudinal muon depolarization rate ($1/T^\mu_1$) under LF = 0.1 T, (c) magnetic fluctuation rate ($\nu^{\rm L}$), and (d) specific-heat divided by $T$ in \lvo. Filled symbols represent data for the high-quality sample (this work), and open symbols are quoted from our previous \msr\ result for powder sample \cite{Koda:04}. Note that (b) and (d) are semi-log plots while (c) is a double-log representation. The cross-hatched region in (c) indicates fluctuation rate deduced from $^7$Li-NMR ($=\nu^{\rm F}$),  dashed line in (c) shows temperature dependence of $\Gamma_{\rm q}=\Gamma({\bf Q}_c)$ inferred from INS \cite{Lee:01}. The dashed/dot-dashed curves in (d) are the corresponding data of $C_p/T$ quoted from Ref.\cite{Kaps:01}.}
\label{nu}
\end{center}
\end{figure}

The Fourier transform of the time-dependent \msr\ spectra $\mathcal{A}_x(t)$ for a transverse field $B_0=0.1$ T, and the parameters deduced from curve-fits for the spectra under $B_0=0.1$ and 0.5 T   are shown in Fig.~\ref{KL}, where the previous data on another set of high quality single-crystalline (sc-) \lvo\ samples under $B_0=1$ T \cite{Koda:05} are also plotted. (For the details on samples and \msr\ experiment, see Supplemental Material \cite{SM}) The muon Knight shift exhibits a divergent behavior that is in marked contrast with $\chi_{\rm bulk}$, and splits into two components, $K_{\mu0}$ and $K_{\mu1}$,  below  $\sim$10$^2$ K with relative signal amplitude of $\sim$10\% or less for $K_{\mu0}$ (see Fig.~\ref{KL}a). In addition,  $1/T^{\mu1}_2$ concomitantly exhibits strong enhancement with increasing $B_0$, where the lineshape at 1 T (not shown) is better represented by assuming further splitting (with $K_{\mu2}<K_{\mu1}$). In contrast, the spectra under a longitudinal field (LF) showed least dependence on the magnitude of $B_0$, as is evident in Fig.~\ref{nu}a (except for the change between 0 and 10 mT corresponding to the quenching of quasistatic nuclear dipolar fields under a weak LF). Because of the relatively small amplitude for $K_{\mu0}$ that tends to decrease with improved sample quality, we attribute this component to an unknown extrinsic phase and focus on the $K_{\mu} \equiv K_{\mu1}$ component below (with $1/T^{\mu}_2\equiv1/T^{\mu1}_2$). 

The spin part of the Knight shift is given by $K_a=A_{a0}\chi'(0,0)/(N_A\mu_B)$,
where the label $a$ is introduced for  distinguishing the cases of \msr\ ($a=\mu$) and NMR ($a=I$), $A_{a0}$ is the ${\bf q}=0$ component of the ${\bf q}$-dependent hyperfine parameter $A_{a{\bf q}}$, $N_A$ is the Avogadro number, and $\mu_B$ is the Bohr magneton.  
A curve-fit analysis of the shift at 0.1 T by the Curie-Weiss law, $K_{\mu}\propto 1/(T+\theta_\mu)$, yields excellent fit (dashed curve in Fig.~\ref{KL}b) with a Weiss temperature as small as $\theta_\mu=1.79(1)$ K. Another excellent fit is obtained by $K_{\mu}\propto-\ln T$ for the data below 10 K  (solid curve in Fig.~\ref{KL}b). For comparison, also plotted in Fig.~\ref{KL}b are the shifts corresponding to $\chi_{\rm bulk}$, $K_{\rm bulk}=A_{\mu0}\chi_{\rm bulk}/N_A\mu_B$, and to $\chi_{\rm s}=\chi({\bf Q}_c,0)$ obtained from INS \cite{Lee:01}, $K_{\chi {\rm q}}=A_{\mu0}\chi_{\rm s}/N_A\mu_B$ [with ${\bf Q}_c$ ($\simeq0.6\:\AA^{-1}$) being the ${\bf q}$ vector characterizing the spatial correlation of the predominant magnetic fluctuation], where $ A_{\mu0}\simeq0.1658$ T/$\mu_B$ was evaluated using the dipolar tensor for the known 16$c$ site \cite{Koda:04,SM}. While $K_{\rm bulk}$ and $K_I$ \cite{Shimizu:12} are in nearly perfect agreement with each other, $K_{\mu}$ exhibits remarkable deviation from these for $T\ll T_K$ with a certain similarity to $K_{\chi {\rm q}}$. It must be noted that $\chi_{\rm bulk}$ exhibits the Curie-Weiss behavior only for $T>T_K$ with a Weiss temperature $\theta_{\rm W}\simeq60$ K, which is much greater than $\theta_\mu$.

It is indicated from Fig.~\ref{KL}c that $1/T^\mu_2$ is strongly enhanced by $B_0$, whereas $1/T^\mu_1$ ($\propto\chi''$) is mostly independent of field (see Fig.~\ref{nu}a) \cite{Koda:04}, suggesting that the linewidth is dominated by the static distribution of shift,  $\Delta K_\mu$ ($\propto\Delta\chi'$), i.e., $(1/T^\mu_2)^2\simeq(1/T^\mu_1)^2+(\Delta K_\mu B_0)^2$.  These features are in remarkable similarity with that observed in the paramagnetic state ($T\ge T_{\rm g}$, the glass temperature) of diluted metallic spin glass systems (e.g., {\it Ag}Mn) \cite{Heffner:82}, suggesting the common origin for the anomalous field-induced inhomogeneity of $\chi'$. Similar anomalies are reported from an earlier $^7$Li-NMR study under a relatively low $B_0$ \cite{Kaps:01}. 

The hyperfine interaction at the muon site ($\overline{3}d$, trigonal) generally consists of two components, i.e., magnetic dipolar interaction ($A^{{\rm L}}_{\mu0}$) with local electrons and transferred hyperfine interaction ($A^{\rm F}_{\mu0}$) with itinerant electrons, where the latter is presumed to yield a minor contribution for muon at interstitial sites. Considering the possibility for these interactions to couple with different components of the local susceptibility, we introduce a phenomenological model in which $\chi$ consists of two parts, i.e., $\chi_{\sigma}({\bf q},\omega)$ with $\sigma={\rm F}$ and ${\rm L}$ \cite{SM}. We employ the conventional Lorentzian form with two components,  
\begin{equation}
\chi_\sigma({\bf q},\omega)=\frac{\chi_{\sigma{\rm s}}}{1+\frac{({\bf q}-{\bf Q}_c)^2}{(\kappa^\sigma)^2}-i\frac{\omega}{\Gamma^\sigma({\bf q})}}=\chi'_\sigma({\bf q},\omega)+i\chi''_\sigma({\bf q},\omega),
\end{equation}
where $\chi_{\sigma{\rm s}}=\chi_\sigma({\bf Q}_c,0)$ is the static susceptibility, $\Gamma^\sigma({\bf q})$ is the magnetic relaxation rate, and $\kappa^\sigma$ is the linewidth. The metallic spin liquid-like behavior described by a local form [$\chi_{\rm loc}(\omega)$] \cite{Sachdev:93,Burdin:02} is presumed to be monitored by the ${\bf q}$-independent parameters. 
The Knight shift is then described as
\begin{equation}
K_a\simeq\frac{1}{N_A\mu_B}\sum_{\sigma={\rm F},{\rm L}}A^{\sigma}_{a0} \chi'_{\sigma{\rm s}},\label{Ks}
\end{equation}
where $\chi'_{\sigma{\rm s}}=\chi'_\sigma(0,0)=\chi_{\sigma{\rm s}}/[1+|{\bf Q}_c|^2/(\kappa^\sigma)^2]$.
Note that the transferred hyperfine interaction dominates for $^7$Li-NMR due to the cubic symmetry of the Li site, so that $K_I\simeq A^{\rm F}_{I0} \chi'_{\rm Fs}/N_A\mu_B$ \cite{Takeda:15}.
Thus, it is interpreted that $K_{\rm Li}$ in Fig.~\ref{KL}b represents the contribution of $\chi'_{\rm Fs}$, and that $K_\mu$ is predominantly determined by $\chi'_{\rm Ls}$ for $T< T_K$ where $K_\mu\simeq A^{\rm L}_{\mu0} \chi'_{\rm Ls}/N_A\mu_B$. The temperature dependence of $K_\mu$ suggests that $\chi_{\rm L}({\bf q},\omega_\mu)$ represents a strongly localized component of the electronic state, which is in line with the dominant role of magnetic dipolar interaction for $A^{\rm L}_{\mu0}$ ($\simeq A_{\mu0}$).

A similar situation is observed for the magnetic relaxation rate among different probes.  
As shown in Fig.~\ref{nu}b, $1/T^\mu_1$ exhibits a tendency of gradual increase and subsequent level-off with decreasing temperature, which is in marked contrast with the case of NMR where $1/T^I_1$ obeys the Korringa relation ($\propto T$) over the relevant temperature region \cite{Shimizu:12,Mahajan:98,Johnston:05}.  The longitudinal depolarization rate is obtained using $\chi''_\sigma({\bf q},\omega)$,
\begin{equation}
\frac{1}{T^{a}_1}\simeq
\frac{k_BT}{N_A\mu_B^2}\sum_{{\bf q},\sigma={\rm F},{\rm L}} \frac{(\gamma_{a}A^\sigma_{a{\bf q}})^2\chi_{\sigma{\rm s}}\Gamma^\sigma({\bf q})}{\omega_{a}^2+[1+({\bf q}-{\bf Q}_c)^2/(\kappa^\sigma)^2]^2[\Gamma^\sigma({\bf q})]^2},\label{Tone}
\end{equation}
 where $\omega_{a}\simeq\gamma_{a}B_0$. To compare the temperature dependence of ${\bf q}$-averaged $\Gamma^\sigma({\bf q})$ ($=\nu^\sigma$) with that deduced from INS ($=\Gamma_{\rm q}$, which is predominantly determined by ${\bf q}\simeq{\bf Q}_c$), we define the ${\bf q}$-averaged quantities,  $2(\delta_a^\sigma)^2$ for $(\gamma_aA^\sigma_{a{\bf q}})^2$, and 
take an approximation of Eq.~(\ref{Tone}) to deduce $\nu^\sigma$ from $1/T^a_1$,
\begin{eqnarray}
\frac{1}{T^I_1T}&\simeq& \frac{k_B\chi_{\rm Fs}}{N_A\mu_B^2}\frac{2(\delta_I^{\rm F})^2}{\nu^{\rm F}},\:\:\label{toneI}\\
\frac{1}{T^\mu_1T}&\simeq& \frac{k_B\chi_{\rm Ls}}{N_A\mu_B^2}\frac{2(\delta_\mu^{\rm L})^2}{\nu^{\rm L}},
\label{toneM}
\end{eqnarray}
where the contribution of $\chi_{\rm Ls}$ to $1/T^I_1$ as well as that of $\chi_{\rm Fs}$ to $1/T^\mu_1$ becomes negligible because $\delta_I^{\rm L}$ ($\propto A_{I0}^{\rm L}$) and $\delta_\mu^{\rm F}$ ($\propto A_{\mu0}^{\rm F}$) are small (as inferred from $K_a$).
(As shown below, the result of the numerical analysis is consistent with the presumption that $\omega_I,\omega_\mu\ll\nu^{\rm F},\nu^{\rm L}$.) In addition, the large difference in the sensitive range of $1/T_1$ between NMR ($1/T^I_1\le 10^0$ s$^{-1}$) and \msr\ ($1/T^\mu_1\ge 10^4$ s$^{-1}$) must be considered \cite{SM}. 

For the NMR part, using the reported hyperfine field for $^7$Li nuclei ($\delta^{\rm F}_I\simeq17.8$--26.9 MHz/$\mu_B$),  $\chi_{\rm bulk}$ for $\chi_{\rm Fs}$, and the Korringa relation, $1/T_1^IT\simeq 2.0$--2.25 s$^{-1}$K$^{-1}$ over a low temperature region $0.5\le T\le 4.2$ K \cite{Johnston:05}, $\nu^{\rm F}$ is estimated from Eq.~(\ref{toneI}) to be $\simeq10^{13}$ s$^{-1}$ (shown as a hatched area in Fig.~\ref{nu}c). This is in reasonable agreement with that expected for the presumed HF quasiparticle state, $\nu_F\simeq2\pi^2k_B^2N_A/3\hbar\gamma=1.7\times10^{13}$ s$^{-1}$ for $\gamma\simeq0.42$ J/mol/K$^2$ observed at 2 K, in  support for the model that $\chi_{\rm F}({\bf q},\omega)$ corresponds to the itinerant part of the electronic state.

For the self-contained evaluation of $\nu^{\rm L}$, we note the relation $\chi_{\rm Ls}/\chi'_{\rm Ls}= 1+|{\bf Q}_c|^2/(\kappa^{\rm L})^2$ between $\chi'_{\rm Ls}$  and $\chi_{\rm Ls}$   in Eqs.~(\ref{Ks}) and (\ref{Tone}).  We can further expect that $\Gamma^{\rm L}({\bf q})/\Gamma^{\rm L}({\bf Q}_c)\simeq 1+({\bf q}-{\bf Q}_c)^2/(\kappa^{\rm L})^2$ for the local spin systems.
Considering that these deviations from unity in proportionality tend to cancel through the ${\bf q}$-average in Eq.~(\ref{Tone}),  we may reasonably assume that substitution of $\chi_{\rm Ls}$ in Eq.~(\ref{toneM}) with $\chi'_{\rm Ls}$ as a better ${\bf q}$-average.
The magnetic relaxation rate $\nu^{\rm L}$ is then deduced from $K_\mu$ and $1/T^\mu_1$ at 0.1 T ($\delta^{\rm L}_\mu\simeq\gamma_\mu A_{\mu0}=141.2$ MHz/$\mu_B$).  As shown in Fig.~\ref{nu}c, the qualitative agreement between $\nu^{\rm L}$ and $\Gamma_{\rm q}$ for $T\ge1.6$ K, in addition to the similarity between $K_\mu$ and $K_{\chi {\rm q}}$, provides evidence that both \msr\ and INS mainly probe $\chi_{\rm L}$. More importantly, $\nu^{\rm L}$ exhibits a general trend of decrease with decreasing temperature (except for a slight retention around 5 K, which we discuss later). For allowing a wider scope for the temperature range, we quote our previous result obtained for a powder specimen cooled down to $\sim$0.02 K, in which one of the two signals showing greater depolarization rate ($\lambda_D$ in Ref.\cite{Koda:04}) turns out to be the relevant component \cite{Kadono:12}. As is evident in Fig.~\ref{nu}b, these two sets of data show smooth overlap with each other, supporting our presumption that $\lambda_D$ represents an intrinsic property. (This owes to the merit of \msr\ as a local probe, with which we can readily distinguish the origin of signals between \lvo\ and other secondary phases.) It is now clear that $\nu^{\rm L}$ [dominated by $\Gamma^{\rm L}({\bf Q}_c)$] exhibits a power law ($\nu^{\rm L}\propto T^\alpha$ with $\alpha\simeq1$) below $\sim$5 K over 2--3 decades in temperature. Such behavior indicates that the Kondo screening is incomplete for the $\chi_{\rm L}$ component, as it has been suggested by absence of $-\ln T$ dependence in $\rho(T)$ \cite{Urano:00}. 

Previously, the deviation of $\chi_{\rm s}$ from the Curie-Weiss law observed by INS for $T<T_K$ (shown by $K_{\chi {\rm q}}$ in Fig.~\ref{KL}b) was interpreted as a sign for the development of the Kondo screening \cite{Lee:01}. However, the recent INS experiment on sc-\lvo\ showed emergence of the second component around ${\bf Q}_c$ with a relatively broader $\kappa$ \cite{Tomiyasu:14} that might have been overlooked as background in the previous experiment, leading to the underestimation of $\chi_{\rm s}$. Thus, the apparent reduction of $\chi_{\rm s}$ may be related to the gradual development of the second component (which was not discernible for $1/T^\mu_1$). 
The weighted average of the reported magnetic relaxation rate $\Gamma$ for these two components is in quantitative agreement with $\nu^{\rm L}\sim$3$\times10^{11}$ s$^{-1}$ at 6 K estimated from $1/T^\mu_1$, supporting the above interpretation. 

We now draw attention to the fact that the critical behavior of $\nu^{\rm L}$ close to that of spin glass (with $T_g=0$) coexists with marginal Fermi liquid behavior, which is indicative of the interesting interrelationship between these phenomena. As shown in Fig.~\ref{nu}d, our earlier data of specific heat ($C_p$) on a sc-\lvo\ sample indicates that $C_p/T$ keeps increasing with decreasing temperature below $T_K$, exhibiting $-\ln T$ dependence for $2\le T\le \sim$10 K without any sign of saturation at 2 K \cite{Matsushita:05}. Such behavior is reported to extend down to $\sim$0.1 K on powder specimen \cite{Kaps:01}, as quoted in Fig.~\ref{nu}d. Considering that $\gamma\simeq C_p/T$ (the quasiparticle mass) for the relevant temperature range, the $-\ln T$ dependence of $C_p/T$ can be regarded as an unambiguous sign for the persistent marginal Fermi liquid character. It is also clearly distinct from the situation that the entropy is entirely carried by the fluctuating local spins \cite{Moriya:95,Hayden:00,Yushankhai:08}, where one would expect $\gamma\propto[\Gamma^{\rm L}({\bf Q}_c)]^{-1}\propto (\nu^{\rm L})^{-1}\propto T^{-\alpha}$ according to the present \msr\ result. 

\begin{figure}
\begin{center}
\includegraphics[width=0.75\linewidth]{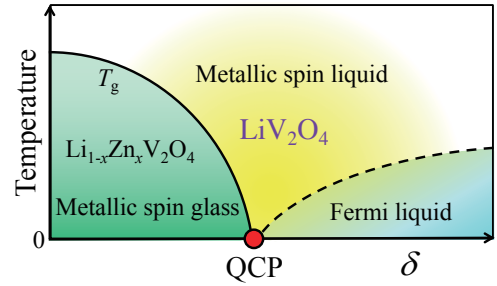}
\caption{(Color online) Schematic phase diagram displaying the quantum critical point (QCP) between metallic spin glass and Fermi liquid states along a control parameter $\delta$. $T_{\rm g}$ is the glass temperature. \lzvo\ is mapped onto the spin glass state, where $0\le x<0.05$ may correspond to the QCP ($T_{\rm g}=0$).}
\label{qpm}
\end{center}
\end{figure}

While the Curie-Weiss behavior of  $\chi_{\rm Ls}$ may be well understood by the self-consistent renormalization (SCR) theory for the AF correlation \cite{Yushankhai:08}, the temperature dependence of $1/T_1^\mu$ is qualitatively different from the predicted behavior of $1/T_1^\mu\propto T^{1/2}$ \cite{Moriya:95}. Moreover, the theory assumes the Fermi liquid state which seems hardly established in \lvo\ over the relevant temperature range due to the insufficient Kondo screening.  Given this situation, we attribute such unusual properties to the putative metallic spin liquid state realized within the $d$-electron band, which is split into $a_{\rm 1g}$ and $e'_{\rm g}$ states by trigonal distortion. As schematically shown in Fig.~\ref{qpm}, \lzvo\ exhibits spin glass behavior with finite $T_{\rm g}$ for $x\ge 0.05$ \cite{Trinkl:00}, and \lvo\ is situated near the endpoint of the metallic spin glass. The field-induced anomalous enhancement of $1/T^\mu_2$ against $1/T^\mu_1$ also provides circumstantial evidence for the microscopic inhomogeneities specific to spin glass.

It might be speculated that $\chi_{\rm L}$ mainly originates from the $a_{\rm 1g}$ band, which becomes almost localized by interorbital electronic correlation \cite{Kusunose:00,Nekrasov:03,Arita:07}. The renormalized band would also serve as a stage for metallic spin liquid-like magnetic excitations.  Regarding local susceptibility for the metallic spin liquid, $\chi_{\rm loc}(\omega)=\chi'_{\rm loc}(\omega)+i\chi''_{\rm loc}(\omega)$, it is predicted that $\chi'_{\rm loc}(0)\propto J^{-1}\ln(J/T)\propto -\ln T$ and  that $1/T_1\propto T\chi''_{\rm loc}(\omega)/\omega\sim1/J\sim const.$ (with $J$ being the mean of the random magnetic exchange coupling energy) \cite{Burdin:02}. The behavior of $K_\mu$ and $1/T^\mu_1$ is in line with these predictions (as the temperature dependence of $K_\mu$ is equally well represented by $-\ln T$). Thus, the ${\bf q}$-averaged $\chi_{\rm L}$ is suggested to represent $\chi_{\rm loc}(\omega)$. Meanwhile, the theory also predicts a crossover to the Fermi liquid state below a characteristic temperature $T^*$ ($<T_K$), which is in contrast with the coexistence of two components inferred for \lvo.

 The behavior of $\chi_{\rm L}$ suggests strong interaction of local spins with the marginal Fermi liquid portion corresponding to $\chi_{\rm F}$ via hybridization and possibly double-exchange interaction (a higher order effect of the Hund coupling). The competition between the Hund coupling and AF correlation via direct exchange interaction may be the origin of the secondary energy scale ($\simeq\theta_\mu$). While this reminds us of the second component of $\chi({\bf q},\omega)$ observed by INS, it cannot be simply attributed to $\chi_{\rm F}$ considering the orders of magnitude difference in the spin fluctuation rate ($\nu^{\rm L}$ vs $\nu^{\rm F}$). 

Finally, we point out that the monotonic decrease of $\nu^{\rm L}$ with decreasing temperatures should entail anomalous response of $\chi_{\rm L}({\bf q},\omega)$ to the external magnetic field at lower temperatures where $\nu^{\rm L}$ becomes comparable with the Zeeman frequency of paramagnetic moments ($\omega_e/B_0=1.761\times10^{11}$ s$^{-1}$/T). Such a matching of the two energy scales will disturb the intrinsic magnetic fluctuation around ${\bf Q}_c$, depending on both temperature (that determines $\nu^{\rm L}$) and the magnitude of $B_0$ to cause $\omega_e\simeq\nu^{\rm L}$. The field-induced increase of $\Delta K$ observed for $\mu$SR at lower temperatures is naturally understood by the blurring of the propagation vector ${\bf Q}_c$ and associated increase of $\kappa^{\rm L}$ that enhances $\chi'_{\rm L}$ via the factor $(Q_c/\kappa^{\rm L})^{-2}$. 

We would like to thank K. Tomiyasu and K. Yamada for helpful discussions on the earlier INS results for \lvo. We also thank the PSI staff for their valuable support concerning the \msr\ experiment.

\end{document}